\documentclass[12pt,twoside]{article}
\usepackage{fleqn,espcrc1}
\usepackage{amssymb,latexsym}
\usepackage{epsfig}

\usepackage{graphicx}
\usepackage[figuresright]{rotating}

\newcommand{\AmS}{{\protect\the\textfont2
  A\kern-.1667em\lower.5ex\hbox{M}\kern-.125emS}}

 \title{Event Texture Search for Critical Fluctuations in Pb+Pb Collisions}
\author
{Mikhail L. KOPYTINE$^j$ \thanks{\protect{
on an unpaid leave from P.N.Lebedev Physical Institute,
Russian Academy of Sciences}}
for The NA44 Collaboration \linebreak
I.Bearden$^{a}$, H.B{\O}ggild$^{a}$, J.Boissevain$^{b}$, L.Conin$^{d}$,
J.Dodd$^{c}$, B.Erazmus$^{d}$, S.Esumi$^{e}$, C.W.Fabjan$^{f}$, 
D.Ferenc$^{g}$, D.E.Fields$^{b}$, A.Franz$^{f}$, J.J.Gaardh{\O}je$^{a}$,
A.G.Hansen$^{a}$, O.Hansen$^{a}$, D.Hardtke$^{i}$, H. van Hecke$^{b}$, 
E.B.Holzer$^{f}$, T.J.Humanic$^i$, P.Hummel$^f$, B.V.Jacak$^j$, R.Jayanti$^i$,
K.Kaimi$^e$, M.Kaneta$^e$, T.Kohama$^e$, M.L.Kopytine$^j$, M.Leltchouk$^c$,
A.Ljubicic, Jr$^g$, B. L{\"o}rstad$^k$, N.Maeda$^e$, L.Martin$^d$, 
A.Medvedev$^c$, M.Murray$^h$, H.Ohnishi$^e$, G.Paic$^f$, S.U.Pandey$^i$,
F.Piuz$^f$, J.Pluta$^d$, V.Polychronakos$^l$, M.Potekhin$^c$, G.Poulard$^f$,
D.Reichhold$^i$, A.Sakaguchi$^e$, J.Schmidt-S{\O}rensen$^k$, J.Simon-Gillo$^b$,
W.Sondheim$^b$, T.Sugitate$^e$, J.P.Sullivan$^b$, Y.Sumi$^e$, W.J.Willis$^c$,
K.L.Wolf$^h$, N.Xu$^b$, D.S.Zachary$^i$
\linebreak
$^a$ Niels Bohr Institute, Denmark;
$^b$ LANL, USA;
$^c$ Columbia U., USA;
$^d$ Nuclear Physics Laboratory of Nantes, France;
$^e$ Hiroshima U., Japan;
$^f$ CERN;
$^g$ Rudjer Boscovic Institute, Croatia;
$^h$ Texas A\&M U., USA;
$^i$ The Ohio State U., USA;
$^j$ SUNY at Stony Brook, USA;
$^k$ U. of Lund, Sweden;
$^l$ BNL, USA.
}

\begin{document}

\maketitle

\begin{abstract}
NA44 uses a 512 channel Si pad array covering 
$1.5 <\eta < 3.3$
to study charged hadron production in 158 A GeV Pb+Pb collisions at
the CERN SPS.  We apply a multiresolution analysis, based on a Discrete Wavelet
Transformation, to probe the texture of particle distributions 
event-by-event,
by simultaneous localization of features in space and scale.  Scanning a
broad range of multiplicities, we look for a possible critical
behaviour in the power spectra of local density fluctuations.  The data
are compared with detailed simulations of detector response, using
heavy ion event generators, and with a reference sample 
created via event mixing.
An upper limit is set on the probability and magnitude of dynamical
fluctuations.
\end{abstract}

\section{Introduction}
The deconfinement/chiral symmetry restoration
 phase transition in ultrarelativistic heavy ion 
collisions is inherently a multiparticle phenomenon. Therefore,
event-by-event analysis of multiparticle hadronic observables is
of paramount interest. We carry out a texture, or local fluctuation,
analysis to determine the correlation/fluctuation content of single
events, and analyze the scale composition of the correlations.

The idea to look at particle distributions in rapidity $y$ to search for
critical behaviour was proposed 
\cite{Scalapino_Sugar,Carruthers_Sarcevic}
based upon a
Ginzburg-Landau type of multihadron production theory
\cite{Scalapino_Sugar},
where a random hadronic field $\phi(y)$ plays the role of an order parameter
in a hadronization transition.
Enhanced large scale correlations of hadrons in $y$ at 
the phase transition
would signal critical fluctuations in the order parameter.
Stephanov and coworkers \cite{tricritical} indicated a second order
QCD phase transition point which should exist
under certain initial conditions, within the reach of today's
experiments.

In our work, 
a power spectrum analysis of event texture in pseudorapidity
$\eta$ and azimuthal angle $\zeta$,
based on a 
Discrete Wavelet Transformation (DWT)\cite{DWT}, is 
performed on a number of large event ensembles sampled according
to their multiplicity, thereby studying the impact parameter dependence
of the observables.
DWT quantifies contributions of different  $\zeta$ and $\eta$ scales
into the event's overall  texture, thus testing the possible 
large scale enhancement.
\begin{figure}[tb]
\hspace{\fill}
\begin{minipage}[t]{75mm}
\epsfysize=4cm
\epsfxsize=7.5cm
\epsfbox{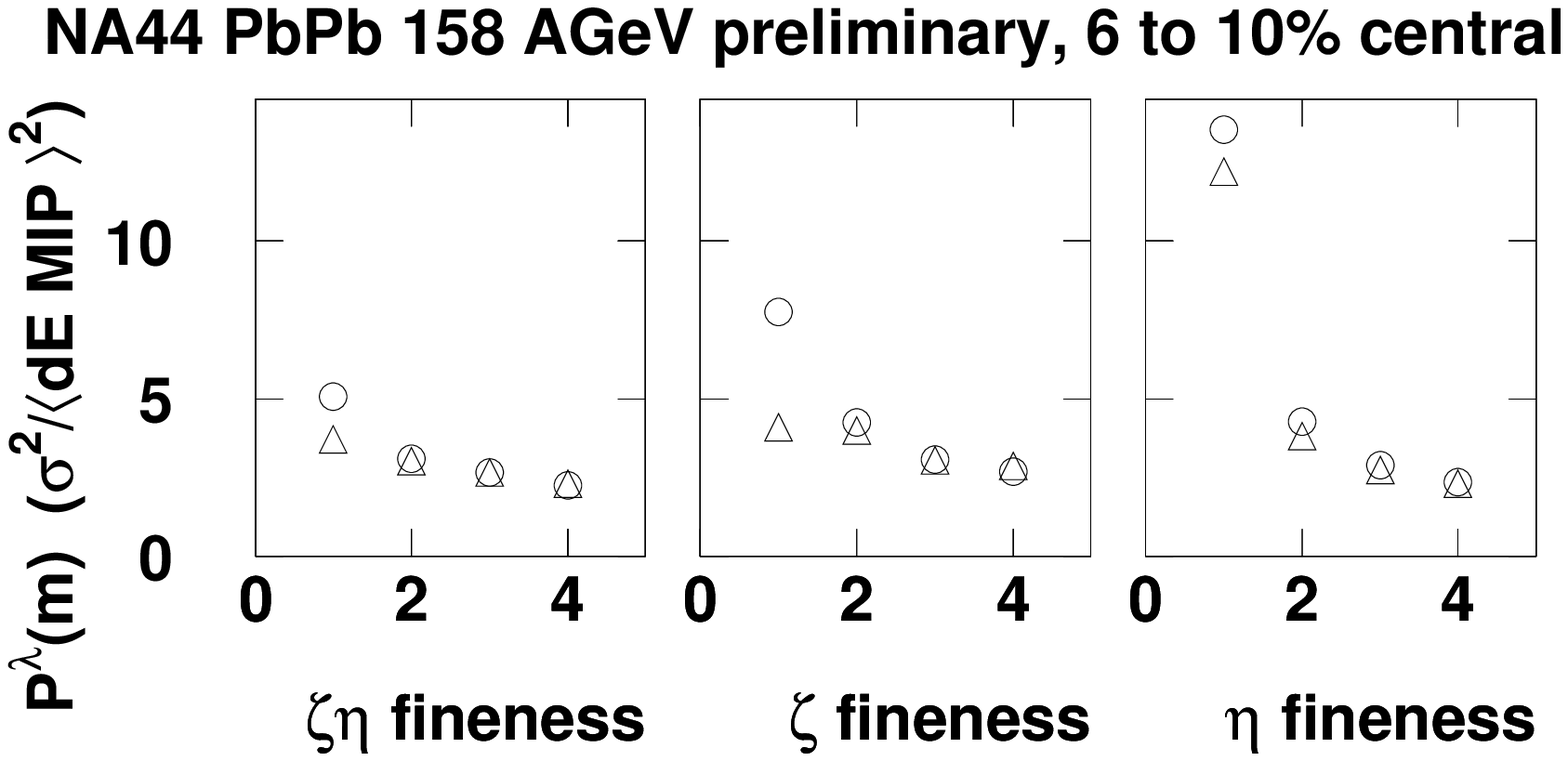}
\caption{Power spectra :
$\bigcirc$ -- true events,
$\bigtriangleup$ -- mixed events.
}
\label{compare}
\end{minipage}
\hspace{\fill}
\begin{minipage}[t]{75mm}
\epsfysize=4cm
\epsfxsize=7.5cm
\epsfbox{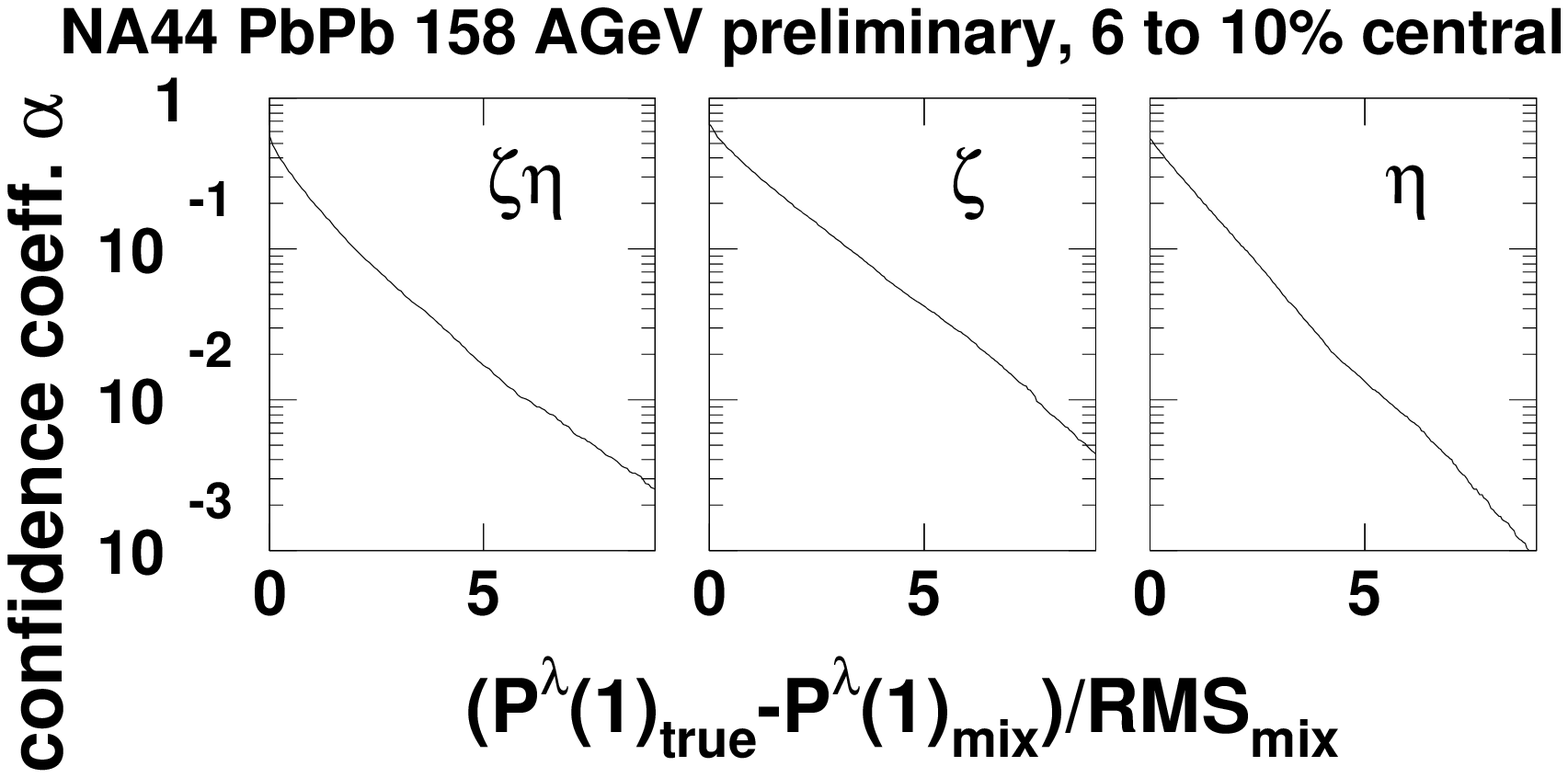}
\caption{Confidence coefficient 
 as a function of the fluctuation strength.
 }
\label{limit}
\end{minipage}
\end{figure}

\section{Experiment and Results}
Ionization energy loss of charged particles was measured in 512
silicon pads, with radial granularity of 16 and azimuthal granularity
of 32.
The pads were read out by AMPLEX 
\cite{AMPLEX}
chips, one chip per sector.
$\delta$-electrons, produced by the $Pb$ beam traversing
the target, were swept away to one side by a dipole magnetic field
($\le 1.6$ Tl).
Only the $\delta$-electron-free side 
was used in the analysis.
Empty target runs were used to measure the background.
Cross-talk in the detector was evaluated off-line.

Track density in an individual event 
$\rho(\zeta,\eta)$ is represented by a 2D array of
the calibrated digitized amplitudes of the channels ( an
\emph{amplitude array}).
This is expanded into a basis of \emph{Haar wavelet}\cite{DWT} 
functions $\Psi^{\lambda}_{m,i,j}(\zeta,\eta)$,
orthogonal with
respect to scale fineness $m$ and location $(i,j)$, with the $\lambda$ index
numbering three modes of texture direction sensitivity --
$\zeta$, $\eta$, and $\zeta\eta$ -- for  azimuthal, pseudorapidity 
and diagonal.
\footnote{
Technically, the 2D basis 
$\Psi^{\lambda}_{m,i,j}(\zeta,\eta) =
 2^{m}\Psi^{\lambda}(2^{m}\zeta-i,2^{m}\eta-j)$
is constructed from 1D functions in the following way:
$ \Psi^\zeta=\psi(\zeta)\phi(\eta),  
\Psi^\eta=\phi(\zeta)\psi(\eta),  
\Psi^{\zeta\eta}=\psi(\zeta)\psi(\eta)$,
where for any variable $x$, the wavelet function 
$\psi(x) =     \{ +1 \mbox{\ for\ } 0\le x<\frac{1}{2}; 
                 -1 \mbox{\ for\ }  \frac{1}{2}\le x<1;
                  0 \mbox{\ otherwise}
              \}$,
and the scaling function
    $\phi(x) = 1$ for $0\le x<1$ and 0 otherwise.            
}
With our binned detector, 
the act of data taking is the first stage of the Haar wavelet transformation
(not true for any other wavelet).

From the squared expansion coefficients a
\emph{power spectrum} is formed:
\begin{equation}
P^\lambda(m) = 
\frac{1}{2^{2m}}\sum_{i,j}\langle \rho,\Psi^\lambda_{m,i,j}\rangle^2 .
\label{eq:P_m}
\end{equation}
       
We used WAILI \cite{WAILI} software to obtain the wavelet expansions.
More experimental and algebraic details can be found in
\cite{ISMD} and the references therein.

\begin{figure}[tb]
\hspace{\fill}
\begin{minipage}[t]{90mm}
\epsfysize=9cm
\epsfbox{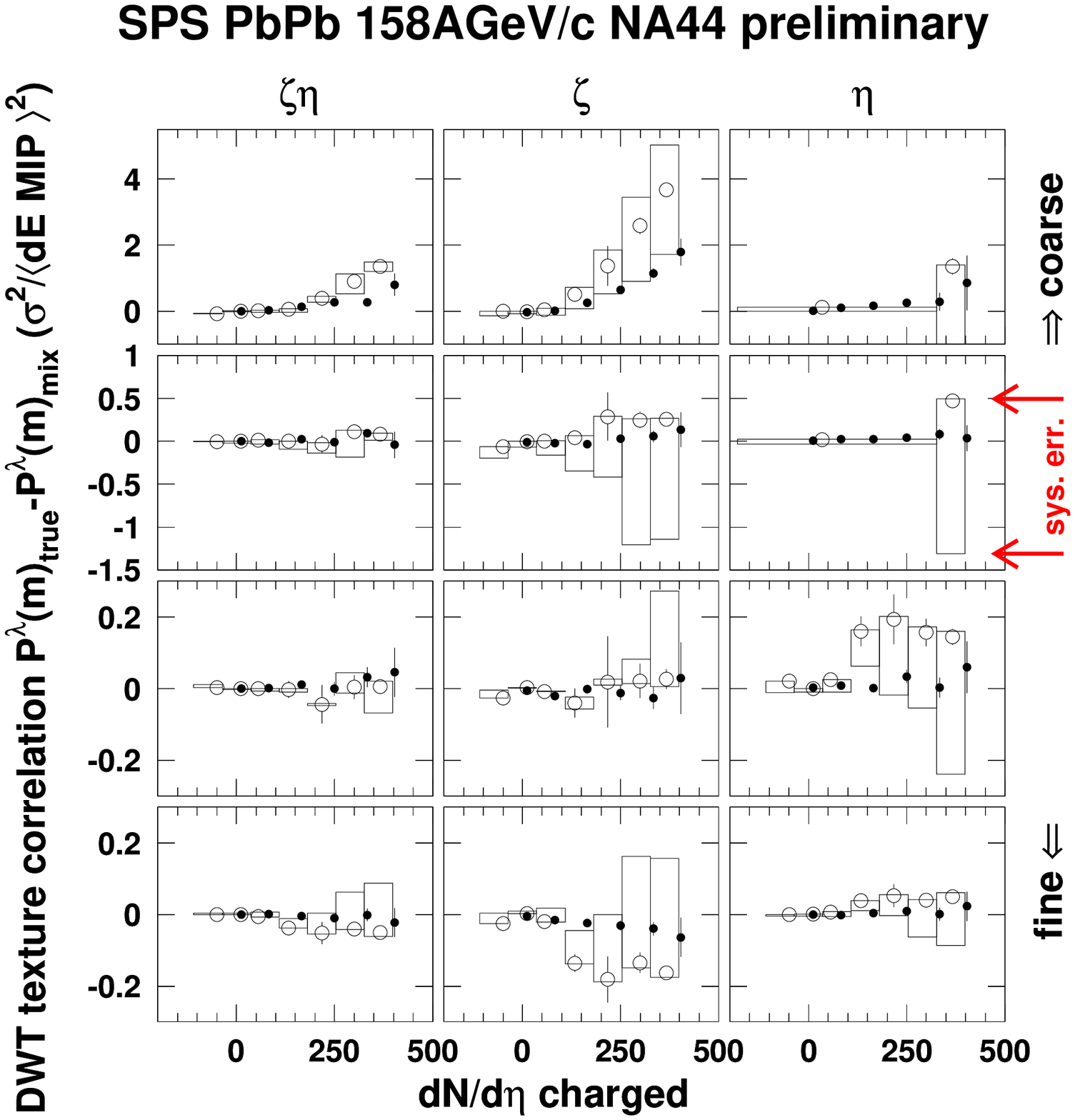}
\caption{
Multiplicity dependence of the texture correlation.
$\bigcirc$ -- the NA44 data, $\bullet$ -- RQMD. 
The boxes show the systematic errors vertically and the boundaries of
the multiplicity bins horizontally; the statistical errors 
are indicated by the vertical bars on the points. The rows correspond
to the scale fineness $m$, the columns -- to the directional mode
$\lambda$ (discussed in the text).
         }
\label{multi_dep}
\end{minipage}
\hspace{\fill}
\begin{minipage}[t]{60mm}
\epsfxsize=6cm
\epsfbox{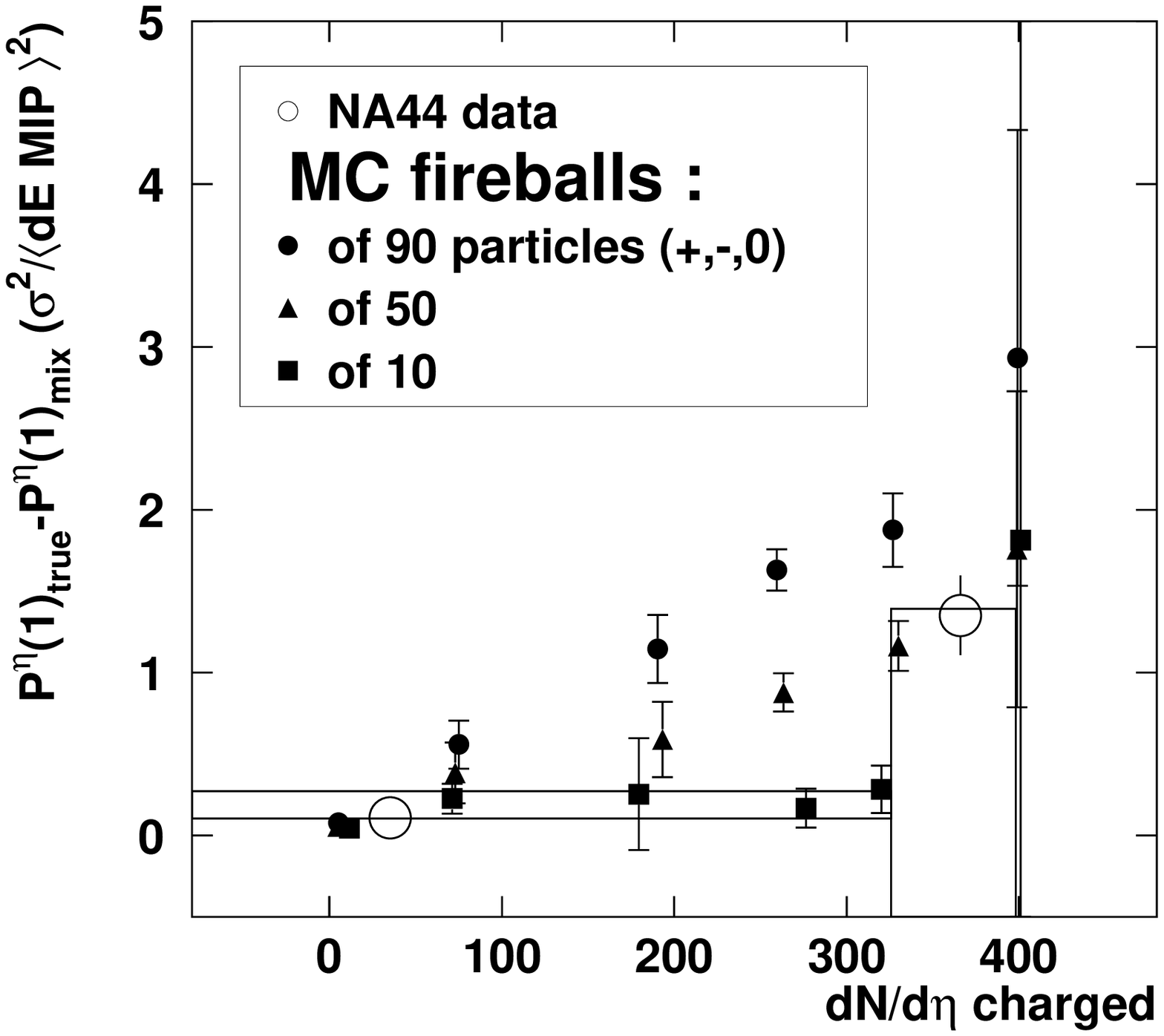}
\caption{
Coarse scale $\eta$ texture correlation in the NA44 data, shown by
$\bigcirc$ (from the top right plot of Figure \ref{multi_dep}),
is compared with that from the multifireball 
event generator for three different fireball sizes.
Detector response is simulated.}
\label{sensitivity}
\end{minipage}
\end{figure}

Figure \ref{compare} 
shows that the power spectra in central Pb+Pb collisions
are indeed enhanced on the coarse scale. 
It is necessary to eliminate
``trivial'' and detector related contributions to this enhancement.
Event mixing is done by 
taking different channels from different events.  
In order to reproduce the electronics cross-talk effects
in the mixed event sample, mixing is done sector-wise, i.e. the
sectors constitute 
the subevents subjected to the event number scrambling.
The mixed events preserve the texture associated with 
detector position offsets,
the inherent $\,dN/\,d\eta$ shape, cross-talk and dead channels.
This is \emph{static} texture as it reproduces
its pattern event after event; 
we are interested in \emph{dynamic} texture.
We reduce sources of  static texture in the 
power spectra by empty target subtraction and by subtraction of the mixed
events power spectra, thus obtaining the \emph{texture correlation}
$P^\lambda(m)_{true} - P^\lambda(m)_{mix}$.
This quantity, normalized to the $RMS$ fluctuation of $P^\lambda(m)_{mix}$,
is used to characterize the relative strength of local fluctuations in an event.
Its distribution for different $\lambda$ is plotted on Figure \ref{limit}
in an integral way, i.e. as an $\alpha(x)$ graph where for every $x$, 
$\alpha$ is the fraction of the distribution above $x$.
The \emph{confidence level}
with which local fluctuations of a strength $x$ can be excluded
is then  $1-\alpha$.
Fluctuations greater than $3\times RMS_{mix}$ are excluded 
in the azimuthal and pseudorapidity modes with 90\% and 95\% confidence,
respectively. 
The monotonic fall of the curve is consistent with absence of abnormal
subsamples in the data.

Figure \ref{multi_dep} summarizes texture correlation data for different
multiplicity events.
The systematic errors were evaluated by removing the $Pb$ target and 
switching magnetic field polarity to expose the analyzed side of the detector 
to $\delta$-electrons, while minimizing nuclear interactions.
Correlations (i.e. deviations of 
$P^\lambda(m)_{true}$ from $P^\lambda(m)_{mix}$) 
in such events measure the remaining systematic uncertainties. 
Thus, this component of the systematic error is signed, 
and the systematic errors are asymmetric. 
The other component (significant only on the coarsest scale) 
is the uncertainty of our knowledge of the beam's geometrical cross-section. 

\section{Discussion and Conclusions}
A Monte Carlo simulation was performed, including known background texture
effects and the subtraction of mixed simulated events.
The finite beam size effect,
irreducible by the subtraction methods, is thereby
taken into account.

The finite beam size effect explains the  rise of RQMD points with
$\,dN/\,d\eta$ in Fig. \ref{multi_dep}.
Finally, the sensitivity of the method is evaluated
 (see Figure \ref{sensitivity}) in MC by 
generating multiple fireball events
with given mean number of particles (charged and neutral), $N_{p}$,
per fireball.
The total $p_T$ of each fireball is 0; its
total $p_Z$ is chosen to simulate longitudinal flow by Lorentz-boosting 
the fireballs along the $Z$ direction,  
keeping the total $\vec{p}$ of an event at 0.
By varying $N_{p}$ per fireball, one varies ``grain coarseness'' of the
event texture in $\eta$.
The data are consistent with clustering of $N_{p}$ per fireball below 50.
 
This novel method of event-by-event analysis,
 applied to the SPS $PbPb$ data, does not reveal any evidence
of critical phenomena.
The authors thank N.Antoniou, I.Dremin, E.Shuryak, M.Stephanov, 
and T.Trainor for illuminating discussions.
This work was supported by the Austrian Fonds zur F{\"o}rderung der
Wissenschaftlichen Forschung;
the Science Research Council of Denmark;
the Japanese Society for the Promotion of Science; the Ministry of
Education, Science and Culture, Japan;  the Science Research Council 
of Sweden; the US Department of Energy and the National Science Foundation.

\end{document}